\documentclass[secnumarabic,
amssymb, nobibnotes, nofootinbib, aps, showpacs,
tightenlines, preprint]{revtex4-1}
\usepackage{amsmath,amssymb,amsthm}
\usepackage{latexsym,graphicx,color,subfigure}
\usepackage[bookmarks=true]{hyperref}
\usepackage{wrapft}
\usepackage{mathrsfs}
\usepackage{slashed}
\usepackage[T1]{fontenc}
\usepackage{amsthm}
\usepackage{amsfonts}
\usepackage{bm}
\usepackage{bbm}
\usepackage{cancel}
\usepackage{pgffor}

\newcommand{\be}{\begin{equation}}
\newcommand{\ee}{\end{equation}} 
\newcommand{\beq}{\begin{eqnarray}}
\newcommand{\eeq}{\end{eqnarray}}

\newcommand{\D}{\mathcal{D}}
\newcommand{\p}{\partial}
\newcommand{\Tr}{{\rm Tr}}

\newcommand{\bea}{\begin{eqnarray}}
\newcommand{\eea}{\end{eqnarray}}
\def\Tr{ \hbox{\rm Tr}}

\def\bra{\langle}
\def\ket{\rangle}
\def\half{\frac{1}{2}}

\def\e{\epsilon}

\def\half{\frac{1}{2}}
\def\dcfl{\Delta_{\textsc{cfl}}}
\def\cfl{\textsc{cfl}}
\def\SU{\rm{SU}}
\def\U{\rm{U}}
\def\bra{\langle}
\def\ket{\rangle}
\def \L{\textsc{l}}
\def \R{\textsc{r}}
\def \B{\textsc{b}}
\def \c{\textsc{c}}
\def \F{\textsc{f}}
\def \M{\textsc{m}}

\begin{document}
\title{{\bf Aharonov-Bohm phase in high density quark matter}}
\author{Chandrasekhar Chatterjee$^{1, 2}$ and Muneto Nitta$^{3}$ }
\email{chandra.chttrj@gmail.com$^{1}$, chandra@phys-h.keio.ac.jp$^{2}$,  
nitta(at)phys-h.keio.ac.jp$^{3}$}
\affiliation{Department of Physics, and Research and Education Center for Natural Sciences,\\ Keio University,Hiyoshi 4-1-1, Yokohama, Kanagawa 223-8521, Japan}
\date{\today}
\begin{abstract}
Stable non-Abelian vortices, 
which are color magnetic flux tubes as well as 
superfluid vortices,
are present in the color-flavor locked
phase of dense quark matter 
with diquark condensations.
We calculate the Aharanov-Bohm phases of 
charged particles, that is, electrons, muons, and  color-flavor locked mesons made 
of tetraquarks around a non-Abelian vortex. 
\end{abstract}
\maketitle
\newpage
\section{Introduction}

The concept of the superfluidity of nuclear matter inside neutron stars was suggested a long time ago by Migdal \cite{Migdal1960}. The mechanism of the Cooper pair formation inside a superconductor due to the electron-phonon interaction can be extended to  the nuclear matter inside a neutron star at sufficiently high density and low temperature, leading  to superfluidity and/or superconductivity 
\cite{Bohr:1958zz, Cooper:1959zz,Baym:1969a}. 
Several astrophysical  observations  indicate that this is likely the case. 
Pulsar glitches \cite{Baym:1969b}, that is the sudden speedup of the rotation frequency of the star,  
were proposed to be explained by 
a sudden unpinning of a large number of 
superfluid vortices in the inner crust of the star 
\cite{Anderson:1975zze};
the observed long-time relaxation  after pulsar glitches
can be explained by 
two components of normal and superfluid neutrons 
\cite{Baym:1969a};
and 
the cooling process of a neutron star was proposed to be
explained by the formation of superconducting or superfluid gaps  \cite{Heinke:2010cr,Page:2010aw}. 

At much higher density, quarks are expected to 
form Cooper pairs to 
show color superconductivity \cite{Alford:1997zt,Alford:1998mk}; 
see Refs.~\cite{Alford:2007xm,Rajagopal:2000wf} 
as a review. 
The two-flavor superconducting (2SC) phase in which 
up and down quarks participate in condensations 
are expected to be realized 
at intermediate density, 
while the 
color-flavor locked (CFL) phase 
in which up, down and strange quarks 
participate in condensations 
is expected to 
be realized 
at asymptotically high density. 
The Ginzburg-Landau free energy in the CFL phase 
was derived in 
Refs.~\cite{Giannakis:2001wz,Iida:2000ha,Iida:2001pg}. 
While magnetic flux tubes are created in type-II metallic superconductor in the presence of magnetic field, color magnetic flux tubes are present stably in the CFL phase 
\cite{Balachandran:2005ev,Nakano:2007dr,Nakano:2008dc,Eto:2009kg}.   
These flux tubes are superfluid vortices created by a rapid rotation of a superconductor; see Ref.~\cite{Eto:2013hoa} as a review. 
This color magnetic flux is a non-Abelian vortex 
carrying collective coordinates 
parametrizing Nambu-Goldstone modes 
${\mathbb C}P^2\simeq 
\SU(3)_{\c+\F}/[\SU(2) \times \U(1)]$ 
localized around the vortex core
that are 
gapless excitations  
propagating along the vortex  
\cite{Eto:2009bh,Eto:2009tr}.
Such vortices will form a vortex lattice 
in rotating color superconductors, 
showing color (anti)ferromagnetism 
\cite{Kobayashi:2013axa}.

The Aharanov-Bohm (AB) effect  \cite{Aharonov:1959fk} 
is a quantum mechanical effect that occurs when a charged particle scatters from a solenoid with  
nonzero magnetic flux inside. 
Outside the solenoid, the field strength is zero everywhere, and the wave function of the 
particle vanishes at the center of the solenoid. 
Nevertheless, 
when a particle goes around the solenoid 
it picks up the phase known as the AB phase, 
leading to a nontrivial differential scattering cross section. 
The AB effect was experimentally confirmed 
\cite{Tonomura:1986} 
and has been studied in various nanomaterials 
in condensed matter physics. Now, the investigation 
is not only limited to materials but is also 
explored in various areas of fundamental physics such as
 cosmology, particle physics, and field theory. 
Vortices or cosmic strings exhibiting the AB effect, 
namely, ``AB cosmic strings,''
were studied extensively \cite{Alford:1988sj}. 
In particular, AB strings feel friction 
as a consequence of  the AB effect 
 \cite{Vilenkin:1991zk,MarchRussell:1991az}.  
AB cosmic strings may give a possible observational 
signature of string theory 
\cite{Polchinski:2005bg,Ookouchi:2013gwa}.
The AB effect around  non-Abelian vortices in 
supersymmetric gauge theory 
was found in Ref.~\cite{Evslin:2013wka}, 
and it has been extended 
\cite{Bolognesi:2015mpa,Bolognesi:2015ida}
to the non-Abelian AB phase 
\cite{Horvathy:1985jr}.
In the context of dense quark matter, 
the AB effect caused by 
a color magnetic flux tube 
was discussed before in the 2SC phase 
\cite{Alford:2010qf},
in which the authors discussed 
scatterings of electrons, muons, and ungapped  quarks via the AB effect.
The friction of vortices and effects on the transport of particles 
were also discussed.
However, color magnetic flux tubes in 
in the 2SC phase are unstable to decay.

In this paper, we investigate the AB effect of 
a color magnetic flux tube (non-Abelian vortex) 
stably existing in 
the CFL phase of dense quark matter. 
In the presence of the electromagnetic interaction, 
a $\U(1)_{\rm em}$ subgroup of the flavor symmetry 
$\SU(3)_{\F}$ is gauged. 
Consequently, an effective potential term on 
the ${\mathbb C}P^2$ space is induced,
resulting in stable and metastable 
vortices with color magnetic fluxes 
correspond to generators 
commuting with $\U(1)_{\rm em}$ \cite{Vinci:2012mc}.
The minimum energy configuration 
is the one found by 
Balachandran, Digal and Matsuura (BDM)
\cite{Balachandran:2005ev}
and the metastable vortices corresponding to 
the ${\mathbb C}P^1$ subspace of which the 
isometry $\SU(2)$ commutes with $\U(1)_{\rm em}$  
(hereafter, we call them  ${\mathbb C}P^1$ vortices).
We calculate the AB phases of charged particles,
 that is, electrons, muons, 
and CFL mesons made of four quarks,
 around a stable BDM vortex or a metastable 
${\mathbb C}P^1$ vortex in the CFL phase. 
Since the  nontrivial AB phase generates frictional force on vortices as denoted above, 
it can affect the conductivity of the vortex particle system and it may create anisotropy in the density of 
particles in the bulk, which remains as a future problem.


The rest of this paper is organized as follows.
In Sec.~\ref{sec:NAvor},
we first give a brief introduction 
of the Ginzburg-Landau(GL) free energy in the CFL phase 
and non-Abelian vortices in the absence 
of the electromagnetic interaction.
Then, in Sec.~\ref{sec:AB}, we introduce 
the electromagnetic interaction by 
gauging a $\U(1)_{\rm em}$ subgroup of the $\SU(3)_{\F}$ flavor symmetry   
and  calculate the AB phases for 
gapless excitation of the CFL phase that 
are scattered by non-Abelian vortices. 
We make a comment on the effect of the strange 
quark mass, in the presence of which 
the AB phase remains nontrivial.

Section~\ref{sec:summary} is devoted to 
a summary and discussion.

 \section{Ginzburg-Landau free energy and non-Abelian vortices in the CFL phase}\label{sec:NAvor}
In this section, we first introduce  the GL description of the CFL phase
and study non-Abelian vortices based on the GL description.

\subsection{Ginzburg-Landau free energy}
The GL description for the order parameter is appropriate at temperatures close to the critical temperature $T_{\rm c}$ for the CFL phase transition. Here the GL order parameters are the diquark condensates $\Phi_{\L/\R}$  defined by
\begin{eqnarray}
&&{\Phi_{\L}}_a^{\it A}  \sim  \e_{abc}\e^{\it ABC} {q_\textsc{l}}_b^{\it B} \mathcal{C}{q_\L}_c^{\it C}, 
\quad 
{\Phi_\R}_a^{\it A}  \sim  \e_{ abc}\e^{\it ABC} {q_\R}_b^{\it B} \mathcal{C}{q_\R}_c^{\it C},
\end{eqnarray}
 where $q_{\L / \R}$ stand for left- and right-handed quarks with ${a, b, c}$ as fundamental color [$\SU(3)_{\c}$] and ${\it A, B, C}$ as fundamental flavor [$\SU(3)_{\L/\R}$] indices. 
 The order parameters $\Phi_{\L/\R}$ transform as a bifundamental representation of color and flavor groups.  It was found that positive parity states are favored compared to the one with negative parity as a ground state. A convenient choice of order parameters for symmetry breaking would be taken as $\Phi_\L = - \Phi_\R \equiv \Phi$. Then, the order parameter $\Phi$ can be regarded as a bifundamental representation of the symmetry group $\U(1)_{\B}\times \SU(3)_{\c} \times \SU(3)_{\F}$.  Here $\U(1)_{\rm{\B}}$ is the global  Abelian transformation of baryon number conservation and the flavor group $ \SU(3)_{\rm{\F}}$ is the diagonal subgroup $\SU(3)_{\L + \R}$ of the total flavor group $ \SU(3)_\L\times \SU(3)_\R$. The GL  free energy can be written in terms of the order parameter $\Phi$ as
  \cite{Giannakis:2001wz,Iida:2000ha,Iida:2001pg}

 \beq
&&\Omega =
\Tr\left[ {1\over 4 \lambda_3} F_{ij}^2 + {\varepsilon_3 \over 2} F_{0i}^2 
+ K_3\D_i \Phi^\dagger \D_i \Phi \right] 
+  \alpha \Tr\left(\Phi^\dagger \Phi \right)
+ \beta_1 \left[\Tr(\Phi^\dagger\Phi)\right]^2 \nonumber\\ 
&&\phantom{xxxxxxxxxxxxxxxxxxxxxxxxxxxxxxxxxxxxxxxx}+\beta_2 \Tr \left[(\Phi^\dagger\Phi)^2\right] + \frac{3\alpha^2}{4(\beta_1+3\beta_2)},
\label{eq:gl}
\eeq
where $i,j=1,2,3$ are indices for space coordinates, $\lambda_{3}$ is a magnetic
permeability, and $\varepsilon_{3}$ is a dielectric constant for gluons.

The GL parameters $\alpha = 4 N(\mu) \log \frac{T}{T_{\rm c}}$, $\beta_1 = \beta_2 = \frac{7\zeta(3)}{8(\pi T_{\rm c})^2}\, N(\mu)\equiv \beta$ and  $K_3 =  \frac{7\zeta(3)}{12(\pi T_{\rm c})^2}N(\mu)$  are obtained from the weak-coupling calculations, which are valid at a sufficiently high density \cite{Iida:2000ha,Giannakis:2001wz}. Here, $\mu$ stands for the quark chemical potential, and we also have taken $\lambda_0 = \epsilon_0 = 
\lambda_3 = \epsilon_3 = 1$. We have introduced the density of state $N(\mu)$ at the Fermi surface $ N(\mu) = \frac{\mu^2}{2\pi^2}.$

\subsection{Non-Abelian vortices}
Let us first briefly review a few salient  features of the non-Abelian vortices in the CFL phase 
in the absence of the electromagnetic interaction.

The covariant derivative and the field strength of gluons are
defined by 
$
\D_\mu \Phi = \p_\mu \Phi - i g_{\rm s} A^a_\mu T^a \Phi, \quad
F_{\mu \nu} = \partial_{\mu}A_{\nu}
-\partial_{\nu}A_{\mu}-ig_{\rm s}[A_{\mu},A_{\nu}]$. 
Here, $\mu$ and $\nu$ are indices for spacetime coordinates and
$g_{\rm s}$ stands for the $\SU(3)_{\c}$ coupling constant. The transformation properties of the field $\Phi$ can be written as
\begin{eqnarray}
 \Phi' = e^{i\theta_\B}\U_\c \Phi \U_\F^{-1}, 
   \quad
 e^{i\theta_\B} \in \U(1)_\B, 
   \quad 
 \U_\c \in \SU(3)_\c,
   \quad 
 \U_\F \in \SU(3)_\F .
\end{eqnarray}
There is a redundancy in the action of the discrete symmetries, and the actual symmetry group is given by
\beq
{\rm G}  =
    \dfrac{\SU(3)_{\c} \times \SU(3)_{\F} \times \U(1)_{\B}}
   {\mathbb{Z}_3 \times \mathbb{Z}_3}.
\label{eq:sym_G}
\eeq
In the ground state
$\bra \Phi \ket = \dcfl {\bf 1_3}$ with  
$\dcfl \equiv \sqrt{-\frac{\alpha}{8\beta}}$,
the full symmetry group $\rm G$ is spontaneously broken down to 
\beq
\rm{H} 
\simeq \dfrac{\SU(3)_{\rm C+F}}{\mathbb{Z}_3}.
\label{eq:H}
\eeq
The order parameter space is 
${\rm G/H} \simeq 
{\SU(3) \times \U(1) \over {\mathbb Z}_3}
=\U(3)$.
 It can be easily noticed that $\pi_1 ({\rm G/H}) = \mathbb Z$. This nonzero fundamental group 
implies the existing vortices. 
Since the broken $\U(1)_{\B}$ is a global symmetry, 
the vortices are global vortices  
or superfluid vortices 
\cite{Balachandran:2005ev}. 
The structure of these vortices can be understood by the orientation and winding of the configuration of the condensed scalar field $\Phi$  far away from the vortex core in perpendicular to  the vortex direction.
We place a vortex along the $z$ direction.

At the large distance $R$ from the vortex core, 
the condensation can have a configuration like
\begin{eqnarray}
\Phi(R, \theta) = 
\dcfl \left(
\begin{array}{ccc}
  e^{i\theta} & 0  & \,0  \\
  0 &  1& \,0 \\
  0 &0 &\, 1  
\end{array}
\right) 
= \dcfl \, \exp {i \left[\frac{\theta}{3}
\left(
\begin{array}{ccc}
 1 & 0 & 0\\
 0 &  1 & 0\\
 0 & 0 & 1
\end{array}
\right) +i  \frac{\theta}{3}
\left(
\begin{array}{ccc}
 2 & 0 & 0 \\
 0 & - 1 & 0\\
 0 & 0& -1
\end{array}
\right)
\right]}.
\label{phiwinding1}
\end{eqnarray}
This can be rewritten as
\begin{eqnarray}\label{phiwinding2}
\Phi(R, \theta) = e^{ig_s\int A\cdot dl} e^{i\frac{\theta}{3}} \Phi(R, 0)
\end{eqnarray}
with $A$ proportional to 
diag.$(2,-1,-1)$.
From Eq.~(\ref{phiwinding2}), 
the minimum energy condition yields
\begin{eqnarray}
D_i \Phi = -i\dfrac{\epsilon_{ij} x_j}{3 r^2} \Phi,\qquad r \rightarrow R 
\end{eqnarray}
at a large distance. From this boundary construction, 
one can write down the ansatz as 
 \begin{eqnarray}
 \label{colorvortexconfig1}
\Phi(r, \theta)  = 
\dcfl\left(
\begin{array}{ccc}
  e^{i\theta}f(r)& 0 & 0 \\
  0 &  g(r) & 0\\
  0 & 0 & g(r)
\end{array}
\right), \, \quad
A_i(r) = - \frac{1}{3g_s} \frac{\epsilon_{ij} x_j}{r^2} [1 - h(r)] \left(
\begin{array}{ccc}
 2 & 0 & 0 \\
 0 & - 1 & 0\\
 0 & 0& -1
\end{array}
\right). 
\end{eqnarray} 
The form of the profiles $f(r)$ and $h(r)$ can be calculated numerically with the boundary condition
\begin{eqnarray}
\label{vortexboundary}
 f(0) = 0, \quad 
\p_r g(r)|_0 = 0, \quad 
h(0) = 1,   \quad f(\infty) = g(\infty) = \dcfl,  \quad 
h(\infty) = 0 .
\end{eqnarray}

The vortex configuration in 
Eq.~(\ref{colorvortexconfig1})  breaks the unbroken 
symmetry $\SU(3)_{\rm{\c+\F}}$ in the ground state 
into a subgroup $\SU(2)\times \U(1)$  
inside the vortex core. 
This breaking results in Nambu-Goldstone modes 
parametrizing a coset space,
    \begin{eqnarray}
\frac{\SU(3)}{\SU(2)\times \U(1)} \simeq {\mathbb C}P^{2}.
\end{eqnarray}
The low-energy excitation and interaction of these zero modes can be calculated by
 the effective ${\mathbb C}P^{2}$ sigma model action 
\cite{Eto:2009bh}.  
Generic solutions on the ${\mathbb C}P^{2}$ space can be found by just applying a global transformation by 
a reducing matrix,
 \begin{eqnarray}
 \label{CP2}
{\U} = 
\frac{1}{\sqrt{X}}\left(
\begin{array}{cc}
 1 & -B^\dagger \\
 B &  X^\half Y^{-\half} 
\end{array}
\right), \quad 
X = 1 + B^\dagger B, \qquad Y = {\bf 1}_3 + BB^\dagger, 
\end{eqnarray}
where $B = \{B_1, B_2\}$ are inhomogeneous 
coordinates of the ${\mathbb C}P^{2}$.
The vortex solution  with a generic orientation and in the regular gauge takes the form
\begin{eqnarray}
 \label{colorvortexconfigcp2}
\Phi(r, \theta)  = 
\dcfl {\U}\left(
\begin{array}{ccc}
  e^{i\theta}f(r)& 0 & 0 \\
  0 &  g(r) & 0\\
  0 & 0 & g(r)
\end{array}
\right){\U^\dagger}, \, 
A_i(r) = -  \frac{\epsilon_{ij}  x_j}{3g_s r^2} [1 - h(r)]  {\U}\left(
\begin{array}{ccc}
 2 & 0 & 0 \\
 0 & - 1 & 0\\
 0 & 0& -1
\end{array}
\right){\U^\dagger}.\nonumber\\
\end{eqnarray}

\section{Aharonov-Bohm phases 
around a non-Abelian vortex}   \label{sec:AB}

As mentioned in Introduction, 
the AB effect \cite{Aharonov:1959fk} 
is a quantum mechanical effect that occurs when a charged particle scatters from a solenoid with  
nonzero magnetic flux inside. 
It leads to the differential scattering cross section
\begin{eqnarray}
\frac{d \sigma}{d\vartheta} = \dfrac{\sin^2(\pi \varphi)}{2\pi k \sin^2(\frac{\vartheta}{2})},\qquad \varphi = \dfrac{q}{ 2 \pi} \times \rm{ Flux }.
\end{eqnarray}
Here $q$ is the electric charge of 
a scattering particle,  $k$ is the momentum perpendicular to the string, and $\vartheta$ is the scattering angle. 
The scattering cross section depends on the flux of the 
solenoid in a nontrivial way.
In the case of a vortex  
carrying a nonquantized flux,
the same thing occurs  
\cite{Alford:1988sj}.
Although 
particles can get inside a vortex core, 
we have the same formula
as far as when we consider paths far from 
the vortex core.

Non-Abelian vortices similar to those in the CFL phase 
in dense QCD 
were found in the CFL phase in supersymmetric gauge theories 
\cite{Auzzi:2003fs,Hanany:2003hp,Eto:2005yh}; 
see Refs.~\cite{Tong:2005un,Eto:2006pg,
Shifman:2007ce,Tong:2008qd} as a review.
When one gauges a $\U(1)$ subgroup of the flavor group, 
non-Abelian vortices become AB strings \cite{Evslin:2013wka}. 
This was extended to non-Abelian gauging \cite{Bolognesi:2015mpa,Bolognesi:2015ida}.
As for a non-Abelian vortex in the CFL phase, 
the AB effect appears once we introduce 
the electro-magnetic interaction [$\U(1)_{\rm em}$] as 
a subgroup of the flavor symmetry group, 
as in the case of supersymmetric theories. So, it would be interesting to determine the value of $\varphi$ for the scattering of particles  that  are the relevant low-energy excitation in the CFL phase. 
In the CFL phase, electrons, muons,  and Nambu-Goldstone bosons, e.g., the CFL mesons, can be considered as fundamental excitations in the  bulk.\footnote{The AB effect can be realized if there exist charged asymptotic states in the bulk of the condensate.  Color charged quasiparticle quarks cannot exist in the bulk freely because of condensation. 
The quark condensate screens color charges in the bulk.
}
Here, we calculate the AB phases of  electrons,  muons  and the CFL mesons present in the bulk.

\subsection{Electromagnetic interactions of non-Abelian vortices}
Here, we introduce $\U(1)_{\rm{em}}$ generator as a part of 
the flavour symmetry $\SU(3)_{\F}$: 
 \begin{eqnarray}
 \label{emcharge}
Q  = \frac{1}{3}
\left(
\begin{array}{ccc}
 2 &  0  & 0  \\
 0 & -1  & 0  \\
 0  &  0 &   -1
\end{array}
\right).
\end{eqnarray}  
Massless symmetry is realized by a linear combination of  color and the $\U(1)_{\rm em}$ subgroup.  
 To see exactly which gauge field remains unbroken, 
let us look at the covariant derivative on the order parameter:
 \begin{eqnarray}
  \D_\mu \Phi 
  = \p_\mu \Phi - i g_s A_\mu  \Phi 
-  i e A^{\rm{em}}_\mu \Phi Q.
\end{eqnarray}
When the order parameter is in a diagonal form 
$\Phi_{\rm diag}$, the covariant derivative 
can be written as  
 \begin{eqnarray}
  \D_\mu \Phi_{\rm diag} 
  = \p_\mu \Phi_{\rm diag} - i\left(g_s A^a_\mu {\rm H}^a  
-  e A^{\rm{em}}_\mu Q\right) \Phi_{\rm diag}.
\end{eqnarray}
Here, we have taken only the color diagonal gauge fields, and ${\rm H}^a = \{\rm{T}^8, \rm{T}^3\}$  are generators of the Cartan subalgebra of the $\SU(3)$ Lie algebra. 
The massive and massless diagonal gauge fields in the bulk 
can be expressed  as (see, e.g.,~Ref.~\cite{Alford:2007xm})
\begin{eqnarray}
&& A^{\M}_\mu = \frac{g_s}{g_M} A^8_\mu - \frac{\eta e}{g_{\M}} A^{\rm{em}}_\mu, 
\quad
A^q_\mu = \frac{\eta e}{g_M}A^8_\mu + \frac{g_s}{g_M}A^{\rm{em}}_\mu ,
\end{eqnarray}
respectively, 
where $\eta = \frac{2}{\sqrt 3}$ and $g_\M^2 = g_s^2 + \eta^2 e^2 $.  
All fields living in the bulk interact with $A^q$ effectively as an effective electromagnetic interaction 
$\tilde\U(1)^{\rm{em}}$ generated by $A^q$. 

The original electromagnetic gauge potential can be written as 
\begin{eqnarray}
\label{Aem}
A^{\rm{em}}_\mu = \frac{g_s}{g_M} A^q_\mu - \frac{\eta e}{g_M} A_\mu^{\M}.
\end{eqnarray}
So, the effective electromagnetic coupling for a particle with charge $q$ becomes 
\begin{eqnarray}
\label{effemcharge}
\frac{q g_s}{\sqrt{g_s^2 + \eta^2 e^2}}.
\end{eqnarray}
Construction of vortices with electromagnetic interaction can be understood from the winding of scalar field and the covariant derivative defined above.

The existence of $ \U(1)_{\rm{em}}$ breaks 
the global $\SU(3)_{\c+ \F}$ invariance to $\SU(2) \times \U(1)$, and consequently 
the ${\mathbb C}P^{2}$  Nambu-Goldstone zero modes become massive, leaving the BDM vortices and $\mathbb{C}P^1$ vortices as (meta)stable configurations\medskip

\underline{BDM vortices}:
In this case, the scalar field configuration at large distance $R$ can be described as
\begin{eqnarray}
\Phi(R, \theta) = 
\dcfl  \left(
\begin{array}{ccc}
  e^{i\theta} &\, 0  & \,\,0  \\
  0 & \, 1& \,\,0 \\
  0 &\,0 &\,\, 1  
\end{array}
\right)  .
\label{phiwindingbdm}
\end{eqnarray}
We can rewrite this in terms of 
a global $\U(1)_B$ rotation added with rotation in 
color and electromagnetic action as
\begin{eqnarray}
\label{phiwindingbdm2}
\Phi(R, \theta) = e^{ig_s\int A\cdot dl}\, e^{i\frac{\theta}{3}} \Phi(R, 0)\,e^{- i e \int  A^{\rm{em}} Q\cdot dl}.
\end{eqnarray}
From this boundary condition, 
one can write down the ansatz as
\begin{eqnarray}
 \label{colorvortexconfig}
 \Phi(r, \theta) =
\dcfl\left(
\begin{array}{ccc}
  e^{i\theta}f(r)& 0 & 0 \\
  0 &  g(r) & 0\\
  0 & 0 & g(r)
\end{array}
\right),  
A^{\rm M}_i(r) {\rm T^8}  = -  \frac{\epsilon_{ij} x_j}{3g_{\textsc{m}}r^2} [1 - h(r)] \left(
\begin{array}{ccc}
 2 & 0 & 0 \\
 0 & - 1 & 0\\
 0 & 0& -1
\end{array}
\right). \label{AMBDM}
\end{eqnarray} 
The form of the profiles $f(r)$ and $h(r)$ can be calculated numerically from the equations of motion with boundary condition in Eq.~(\ref{vortexboundary}) \cite{Balachandran:2005ev}.

\underline{${\mathbb C}P^{1}$ vortices}:
A ${\mathbb C}P^{1}$ sector 
at $|B| \rightarrow \infty$ solutions of 
Eq.~(\ref{colorvortexconfigcp2}) remains gapless \cite{Vinci:2012mc} 
even in the presence of the electromagnetic interaction.
The vortex configurations can be written as 
 \begin{eqnarray}
\Phi(r, \theta) & = & 
\dcfl\left(
\begin{array}{ccc}
g(r)& 0  & 0  \\
  0& e^{i\theta}f(r)& 0\\
  0 & 0 &  g(r) 
\end{array}
\right) , \nonumber \\
\label{AMCP+}
A^{\rm M}_i(r) \rm T^8 &=&  \frac{1}{6 g_{\rm M}} \frac{\epsilon_{ij x_j}}{r^2} [1 - h(r)] \left(
\begin{array}{ccc}
 2 & 0 & 0 \\
 0 & - 1 & 0\\
 0 & 0& -1
\end{array}
\right) , \nonumber \\
A^3_i(r) \rm T^3 &=& - \frac{1}{2g_s} \frac{\epsilon_{ij x_j}}{r^2} [1 - h(r)] \left(
\begin{array}{ccc}
 0\, &\, 0 & 0 \\
 0\, & \, 1 & 0\\
 0\, &\, 0& -1
\end{array}
\right). 
 \label{colorvortexconfigCP+}
\end{eqnarray} 
It is clear from Eq.~(\ref{colorvortexconfigCP+}) that the existence of this vortex spontaneously breaks 
the global $\U(2)$ invariance acting on 
the lower-right 2 by 2 block, and this breaking generates
${\mathbb C}P^{1}$ Nambu-Goldstone modes. 
It is important to note that the configuration of $A^\M_\mu$ 
in Eq.~(\ref{colorvortexconfigCP+})
has a factor $- \half$ compared with 
that in Eq.~(\ref{AMBDM}).

\subsection{Aharonov-Bohm phases of electrons and muons}
The AB scattering of electrons or muons in the CFL phase can be understood by writing the Dirac equation in the vortex background,
\begin{eqnarray}
\left({\slashed\partial} -  i e \slashed A^{\rm em} + i M_{e/\mu_e}\right)\psi_{e/ \mu_e} = 0, 
\end{eqnarray}
where $\psi_{e/ \mu_e}$ are the Dirac fields for electrons and muons with masses $M_{e/\mu_e}$.
  Using Eq.~(\ref{Aem}) we can write 
\begin{eqnarray}
\left(\slashed{\partial} - i  \frac{e g_s}{g_\M} \slashed A^q_\mu + i \frac{\eta e^2}{g_\M} \slashed A_\mu^{\M} + i M_{e/\mu_e} \right)\psi_{e/\mu_e} = 0 .
\end{eqnarray}
The second term is just the coulomb term with effective charge $\frac{e g_s}{g_\M}$, but for the AB scattering the last term would be important. 
This can be understood in another way. 
The AB phase for electrons or muons can be  defined as
\begin{eqnarray}
\label{eAB}
\varphi_{e/\mu_e} = - \frac{e}{ 2 \pi} \oint A^{\rm{em}}\cdot dl. 
\end{eqnarray}
So, according to Eq.~(\ref{Aem}), we may 
calculate the above integral as 
\begin{eqnarray}
\varphi_{e/\mu_e} =  \frac{\eta e^2}{2 \pi g_\M} \oint A^{\M}\cdot dl.
\end{eqnarray}
Here, we have used the fact that 
\begin{eqnarray}
\oint A^q \cdot dl = 0. 
\end{eqnarray}
$A_i^{\M}$ can be determined for the BDM case and for the ${\mathbb C}P^1$ case from Eqs.~(\ref{AMBDM}) and 
(\ref{AMCP+}).   
So the AB
phase around a  BDM vortex can be calculated as 
\begin{eqnarray}
\label{ABbdm}
\varphi^{\textsc{bdm}}_{e/\mu_e} =  \frac{\eta e^2}{2 \pi g_\M} \oint A^{\M}\cdot dl &=& \frac{\eta e^2}{2 \pi g_\M} \times \frac{\eta 2 \pi}{ g_\M}
=  \frac{2 e^2}{3 g_s^2 + 2 e^2},
\end{eqnarray}
while the AB phase $\varphi^{\textsc{cp}^1}_{e/\mu_e}$ around a ${\mathbb C}P^1$ vortex can be determined as
\begin{eqnarray}
\label{ABcp1}
\varphi^{\mathbb{C}P^1}_{e/\mu_e} =  \frac{\eta e^2}{2 \pi g_\M} \oint A^{\M}\cdot dl &=& - \frac{\eta e^2}{2 \pi g_\M} \times \frac{\eta \pi}{ 2 g_\M}
=-  \frac{e^2}{3 g_s^2 + 2 e^2}.
\end{eqnarray}

\subsection{Aharonov-Bohm phases of CFL mesons}
At high density, the chiral symmetry breaking generates Nambu-Goldstone bosons, known as the CFL mesons. The CFL mesons can be expressed using a composite operator of diquark field as
 \begin{eqnarray}
\Sigma_{\cfl}^{AB} = {\Phi^\dagger}^{\L}_{Aa}\Phi^{\R}_{aB}  .
\end{eqnarray}
Here, $a$ and $A$ and $B$ are the color and flavor indices, 
respectively.  
In terms of quarks, the CFL mesons can be expressed as  \cite{Schafer:2000et} 
\begin{eqnarray}
\Sigma^{AB}_{\cfl} 
\sim \e^{ACD} \e^{BEF} 
 {{\bar q}_{\L (a}}{}^{C} { \bar {q}_{\L b)}}^{D}
 {q_{\R (a}}{}^{E}{q_{\R b)}}{}^{F},
\end{eqnarray} 
where $( ...)$ denotes the  antisymmetrization of indices.
The electromagnetic $\U(1)_{\rm em}$ group acts on this 
operator as 
\begin{eqnarray}
\Sigma'_{\cfl} = e^{i e Q \alpha} \Sigma_{\cfl}e^{- ie Q \alpha},
\end{eqnarray}
where $Q$ is defined by Eq.~(\ref{emcharge}). So, the charge can be measured by computing the simple commutator $[Q, \Sigma_{\cfl}]$. 
As we know, $Q$ is basically the $\textsc T^8$ generator of $\SU(3)$ and $\Sigma_{\cfl}$ could also be expanded in $\SU(3)$ generators. There are only four components
of $\Sigma_{\cfl}$ that do not commute with $Q$, which can be written as
\begin{eqnarray}
\left(
\begin{array}{ccc}
 0 & {\Sigma^1}^+_{\cfl}   & {\Sigma^2}^+_{\cfl}   \\
 {\Sigma^1}^-_{\cfl} &   0 & 0 \\
 {\Sigma^2}^-_{\cfl} &   0 & 0
\end{array}
\right).
\end{eqnarray}
 So the charges of $\Sigma_{\cfl}$ mesons can be determined as 
\begin{eqnarray}
q = \{0, \pm e\}.
\end{eqnarray}
In terms of quarks, the charged CFL mesons are
\begin{eqnarray}
&&{\Sigma^1}^+_{\cfl} = \Sigma^{\rm 12}_{\cfl} \sim {\bar {d}_{\L}}{  \bar {s}_{\L}}{s_{\R}}{u_{\R}}, \qquad {\Sigma^1}^-_{\cfl} = \Sigma^{\rm 21}_{\cfl} \sim {  \bar {u}_{\L}}{\bar {s}_{\L}} {s_{\R}}{d_{\R}} \nonumber \\
&&{\Sigma^2}^+_{\cfl} = \Sigma^{\rm 13}_{\cfl} \sim {  \bar {s}_{\L}}{\bar {d}_{\L}} {d_{\R}}{u_{\R}} ,  \qquad {\Sigma^2}^-_{\cfl} = \Sigma^{\rm 31}_{\cfl} \sim {\bar {u}_{\L}} {  \bar {d}_{\L}}{d_{\R}}{s_{\R}}.
\end{eqnarray}
 
The AB phases $\varphi_{\cfl}$ for charged CFL mesons $\Sigma^{\pm i}$ can be expressed by using Eqs.~(\ref{ABbdm}) and (\ref{ABcp1}).
The AB phases for ${\mathbb C}P^1$ vortices ($\varphi^{{\mathbb C}P^1}_{\cfl}$) and  BDM vortices ($\varphi^{\rm{BDM}}_{\cfl}$) can be calculated as 
\begin{eqnarray}
\varphi^{\rm{BDM}}_{\cfl} 
=  \pm \frac{2 e^2}{3 g_s^2 + 2 e^2}, \qquad
\varphi^{{\mathbb C}P^1}_{\cfl} 
= \mp  \frac{e^2}{3 g_s^2 + 2 e^2}.
\end{eqnarray}

\subsection{Strange quark mass}

The importance  of $\mathbb{C}P^1$ vortices can be understood if we study the vortices at an intermediate density regime, which is more relevant in the core of neutron star. In this case, the mass of the strange quark($m_s$) becomes admissible and cannot be neglected. The potential in Eq.~(\ref{eq:gl}) has to be changed by terms like 
$\Tr[\Phi^\dagger\{(\alpha +\frac{2 \e}{3} ){\bf 1} + \e \, T^3 \}\Phi]$, where $\e \propto  m_s^2$. This potential would generate instabilities in the effective theory of 
non-Abelian vortices. 
The general  $\mathbb{C}P^2$ vortices would decay radically with lifetime of order $10^{-21}$ sec, as  estimated in Ref.~\cite{Eto:2009tr} for the case in which $\mu \sim 500$ MeV, $\Delta \sim 10$ MeV and $m_s \sim 150$ Mev. Only one type of $\mathbb{C}P^1$ vortex corresponding to a single point $(0, 1, 0)$ in full  $\mathbb{C}P^2$ moduli space would survive. 
So, only one of ${\mathbb C}P^1$ vortices  becomes  a stable vortex in the presence of the strange quark mass  \cite{Eto:2009tr} as mentioned above. Therefore, in such a situation all vortices have the AB phase $\varphi^{\mathbb{C}P^1}_{e/\mu_e}$.

\section{Summary and discussion} \label{sec:summary}
We have calculated the phases of the AB scattering of the gapless fundamental excitations in 
the CFL phase of dense quark matter 
and have found nontrivial AB phases due to the scattering of electrons, muons, and CFL mesons with vortices.
The nontrivial AB phases arise 
because the flux due to the $\U(1)_{\rm em}$ gauge field shares a fraction of the total magnetic flux present inside vortices and the existence of particles with electric charges present in the bulk of the dense QCD medium as gapless excitations. In the absence of the electromagnetism, 
non-Abelian vortices are degenerate and can be rotated in the ${\mathbb C}P^2$ moduli space resulting in
the effective action 
written as the ${\mathbb C}P^2$ sigma model. 
The presence of $\U(1)_{\rm em}$ as a subgroup of flavor breaks the $\SU(3)$ global invariance and generates a potential in the ${\mathbb C}P^2$ model. In this case, only stable vortices are those for which the color gauge field direction and $\U(1)_{\rm em}$ directions are parallel.
 We have found a missmatch in the AB phases between scattering with BDM vortices (corresponding to 
the $B = 0$ point in the ${\mathbb C}P^2$ moduli space) and ${\mathbb C}P^1$ vortices (corresponding to 
the $B = \infty$ submanifold
on the ${\mathbb C}P^2$ moduli space).  This missmatch arises because of the fact that 
the orientation of color flux to the ${\mathbb C}P^1$  direction changes the fraction of the flux shared by color magnetic field. So, the fraction of electromagnetic flux changes automatically.

The AB scattering off non-Abelian vortices present in the CFL phase is important property of the particles present in the bulk of the CFL phase medium, 
as was discussed for unstable vortices 
in the 2SC phase \cite{Alford:2010qf}.  
We will discuss transportation properties of particles, 
the friction of vortices in the CFL phase, 
and possible implications on physics of neutron stars.

In this paper, we have discussed the AB scattering 
of a single vortex. In the CFL phase under rotation, 
a vortex lattice will be formed. 
The interaction of the electromagnetic field 
with a vortex lattice was discussed in Ref.~\cite{Hirono:2012ki},
showing that the lattice behaves as a polarizer.
The AB scattering of charged particles inside a vortex 
lattice should be an interesting future direction.

We have discussed the AB scattering of charged particles due to the electromagnetic field in the presence of 
a non-Abelian vortex. 
Non-Abelian vortices are 
color magnetic fluxes having non-Abelian fluxes too. 
Since gluons are massive, the AB phase is usually 
thought to be absent, but  
they may give a global analog of 
the AB phase. 
Colored particles in the nontrivial representation of 
the color $\SU(3)_{\c}$ group may have such a phase.
The interaction of quasiquarks 
with a non-Abelian vortex \cite{Yasui:2010yw,Fujiwara:2011za, Chatterjee:2016ykq}
and the interaction of gluons with a vortex \cite{Hirono:2010gq}
 were studied before. 
The presence or absence of the (global) AB phases 
of these colored particles should be clarified.

\section*{Acknowledgments}
This work is supported by the MEXT-Supported Program for the Strategic
Research Foundation at Private Universities ``Topological Science''
(Grant No.~S1511006).
The work of M.~N.~is supported in part by a Grant-in-Aid for
Scientific Research on Innovative Areas ``Topological Materials
Science'' (KAKENHI Grant No.~15H05855) and ``Nuclear Matter in Neutron
Stars Investigated by Experiments and Astronomical Observations''
(KAKENHI Grant No.~15H00841) from the the Ministry of Education,
Culture, Sports, Science (MEXT) of Japan. The work of M.~N.~is also
supported in part by the Japan Society for the Promotion of Science
(JSPS) Grant-in-Aid for Scientific Research (KAKENHI Grant
No.~25400268).


\end{document}